# Coherent control of electron-nuclear states of rare-earth ions in crystals using radio-frequency and microwave radiation

M. R. Gafurov, G.V. Mamin, E.I. Baibekov, I. N. Kurkin,

F.F. Murzakhanov, S.B. Orlinskii

Kazan Federal University, Kazan, Russian Federation, marat.gafurov@kpfu.ru

Electron and/or nuclear spin states are well-known for their use as potential qubits for quantum computing [1]. Single-qubit operation on an electron spin-1/2 qubit can be performed by using microwave (mw) pulse with frequency matching the energy of the transition between the two spin states. Double resonance techniques allow one to manipulate more than two quantum states simultaneously during the pulse sequence. Thus, certain two-qubit operations can be implemented [2].

We have performed electron-nuclear double resonance (ENDOR) and double electron-electron resonance (DEER) manipulations of electronic and nuclear spin states of trivalent gadolinium impurity ions in $CaWO_4$ host crystal. The sample of $CaWO_4$:$Gd^{3+}$ (0.01 at. %) single crystal was grown in Kazan Federal University by the Czochralski method. Experimental studies were done by using helium flow cryostats at X-band Bruker Elexsys E580 spectrometer (T = 5-25 K, microwave frequency $\nu \approx 9.7$ GHz). Pulsed electron paramagnetic resonance spectrum of $Gd^{3+}$ ion in $CaWO_4$ host crystal (magnetic field **B** || $c$ axis) is shown in Fig. 1.

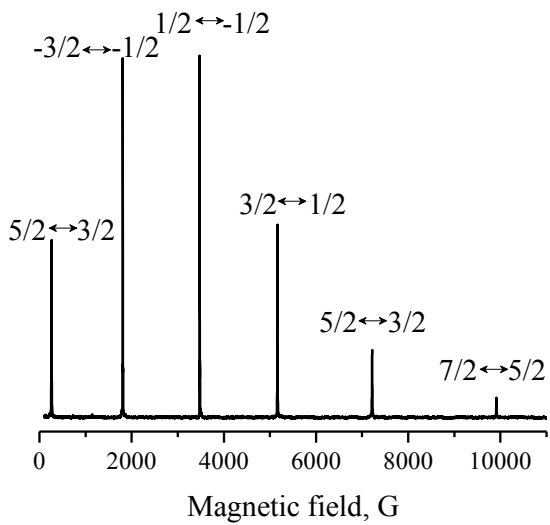

**Fig. 1.** Pulsed EPR of $Gd^{3+}$ ion in $CaWO_4$ crystal, **B** || $c$ axis

It consists of six resonance lines corresponding to various $S_{z1} \to S_{z2}$ transitions between the states of the ground spin-7/2 multiplet. Containing 8 lowest-energy spin states ($S_z = -7/2 \div 7/2$), $Gd^{3+}$ ion can be viewed as model three-qubit system. Single-qubit manipulations in this system have been demonstrated in our previous work [3]. There, mw pulse excited one of the six transitions (chosen by appropriately tuned magnetic field), and the resulting Rabi oscillations were recorded in the time domain of the mw pulse.

Since the impurity gadolinium ions contain odd $^{155}Gd$ and $^{157}Gd$ isotopes with nuclear spin $I = 3/2$ (natural abundance 14.8% and 15.6%, respectively), one can use pulsed ENDOR in order to manipulate both the electron and nuclear spin states of these isotopes simultaneously. The total basis of 8·4 = 32 electro-nuclear states would expand the possible quantum algorithm to 5-qubit operations. In order to demonstrate the control of both the electron and nuclear spin states simultaneously, we used Davies ENDOR pulse sequence [4]: an initial mw π pulse is followed by the radiofrequency (rf) π pulse, then the remaining electron state is probed by the mw two-pulse spin-echo sequence. If the rf pulse frequency matches the nuclear transition, the final echo intensity changes. An example of ENDOR spectrum corresponding to 5/2→3/2 electronic transition near $B = 244$ G swept in the range 0-50 MHz of the rf pulse is shown in Fig. 2. Here, four peaks corresponding to certain electronuclear transitions (two pairs corresponding to $^{155}Gd$ and $^{157}Gd$ nuclear spin flips) are present. The positions of the peaks agree well with our theoretical estimations.

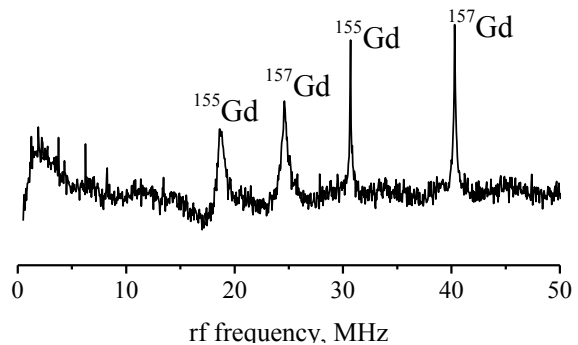

**Fig. 2.** Pulsed Davies ENDOR of $Gd^{3+}$ in $CaWO_4$ crystal, $B = 244$ G is directed along the crystal $c$ axis, $T = 20$ K

Since the positions of the lines in Fig. 1 are orientation-dependent [5], one can choose an appropriate direction of **B** with respect to the crystallographic axes, so that two electronic transitions lie within the 800 MHz bandwidth of the two-frequency DEER equipment. EPR spectrum in certain orientation (13 degrees between **B** and $c$ axis) is shown in Fig. 3. There, two high-field transitions, 5/2→3/2 and 7/2→5/2, involving three electronic levels, are drawn together, the corresponding resonance fields differ by ≈ 100 G. Then, the following two-frequency pulse sequence has been applied: $\pi/2(\nu_1) - \pi/2(\nu_1) - \pi(\nu_2) - \pi/2(\nu_1)$, where the indices in brackets denote the mw pulse frequency. There, $\nu_1 = 9.74$ GHz corresponded to the resonant transition 5/2→3/2 at 7210 G, while $\nu_2$ was swept in the range 9.3-10 GHz. When $\nu_2$ did not match any transition, a usual three-pulse stimulated echo for 5/2→3/2 transition was recorded. However, when $\nu_2$ matched either 5/2→3/2 or 7/2→5/2 transition energy, the third (flipping) π pulse inversed the populations of the two involved spin states, so that the echo intensity was decreased.

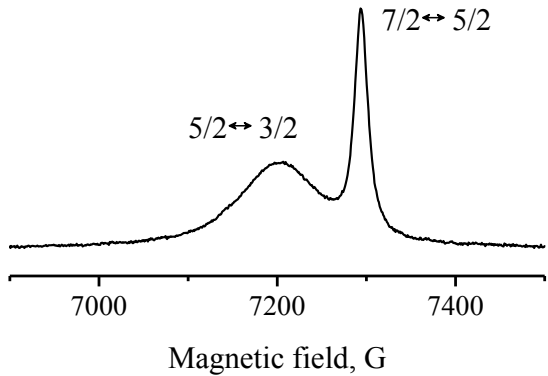

**Fig. 3.** A part of pulsed EPR spectrum of Gd:CaWO$_4$ when **B** is directed at an angle of 13 degrees with respect to crystal *c* axis

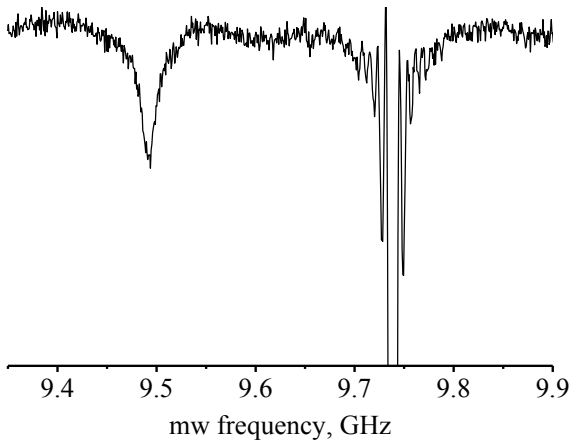

**Fig. 4.** DEER of Gd$^{3+}$ in CaWO$_4$ crystal, $B = 7210$ G, $T = 20$ K, see text for full description

The coherence loss associated with the third pulse resulted in two dips in the echo intensity: at $\nu_2 = \nu_1$ (original 5/2→3/2 transition), and at $\nu_2 = 9.49$ GHz (7/2→5/2 transition), see Fig. 4. An estimated magnitude of the observed hole burnt is ~ 10% of the one at $\nu_2 = \nu_1$, this figure is attributed to the inhomogeneous broadening of the resonance lines. In order to measure the corresponding coherence time of the three-state ensemble (a qutrit), we have recorded the transient nutation of the 7/2→5/2 coherence. There, the duration of the third $\nu_2$ pulse resonant with 7/2→5/2 transition was varied in the range 0-4 μs, while the duration of the π($\nu_2$) pulse was ≈ 180 ns. The resultant echo intensity vs. the pulse duration is shown in Fig. 5.

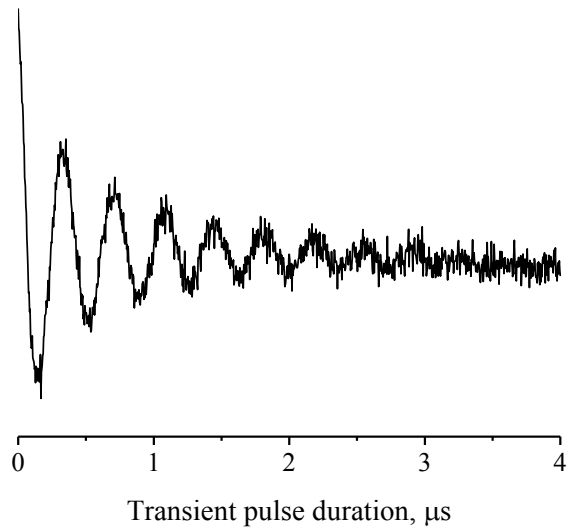

**Fig. 5.** Oscillations of the 5/2→3/2 transition coherence implemented by pulsed DEER, see text for full description

Finally, we have demonstrated electron-electron and electron-nuclear spin manipulations of Gd$^{3+}$ ion in CaWO$_4$ crystal. The results suggest that the studied system is perspective for multiqubit implementation in quantum computing.

This work was financially supported by the Russian Science Foundation (Project no. 17-72-20053).

The work was presented at the 3$^{rd}$ International Conference "Terahertz and Microwave Radiation: Generation, Detection and Applications" (TERA-2018), Nizhny Novgorod, Russia, October 22-25, 2018.